\documentclass[floatfix,preprint, onecolumn, showpacs,nofootinbib]{revtex4-1}
\usepackage{natbib}
\usepackage{times}
\usepackage{amssymb,amsbsy,amsmath,amsfonts}
\usepackage{graphicx}

\usepackage{float}
\usepackage{rotating}
\usepackage{slashed}
\usepackage{color}
\def\slasheps{\varepsilon \!\!\! \slash}
\begin{document}
\newcommand{\ben}[1]{{\color{blue} #1}}
\newcommand{\jorge}[1]{{\color{red} #1}}

\title{Weak decays of unstable $b$-mesons}

\author{B.~Grinstein$^1$ and J.~Martin Camalich$^{1,2}$}
\affiliation{
$^1$Dept. Physics, University of California, San Diego, 9500 Gilman Drive, La Jolla, CA 92093-0319, USA\\
$^2$PRISMA Cluster of Excellence Institut f\"ur Kernphysik, Johannes Gutenberg-Universit\"at Mainz, 55128 Mainz, Germany}

\begin{abstract}

We investigate the decays of the excited $(b\bar q)$ mesons as probes of the short-distance structure of
the weak $\Delta B=1$ transitions. These states are unstable under the electromagnetic or
strong interactions although their widths are typically suppressed by phase space. 
As compared to the pseudoscalar $B$ meson, the purely leptonic decays of the vector $B^*$ are not chirally suppressed and are sensitive to
different combinations of the underlying weak effective operators.
An interesting example is $B^*_s\to \ell^+\ell^-$, which has a rate that can be accurately predicted
in the standard model. The branching fraction is $\mathcal{B}\sim10^{-11}$, 
irrespective of the lepton flavor and where the main uncertainty stems from the unmeasured and theoretically
not-well known $B_s^*$ width. We discuss the prospects for producing this decay mode at the LHC and
explore the possibility of measuring the $B_s^*\to\ell\ell$ amplitude, instead, through scattering 
experiments at the $B_s^*$ resonance peak. Finally we also discuss the charged-current $B_{u,c}^*\to\ell\nu$
decay which can provide complementary information on the $b\to u\ell\nu$ and $b\to c\ell\nu$ transitions.   

\end{abstract}

\maketitle

\section{Introduction}

Heavy-light systems like the $(b\bar q)$ mesons have a rich spectrum of excited states~\cite{Godfrey:1986wj,Isgur:1991wq,Eichten:1993ub,Ebert:1997nk}.
These mesons are unstable under electromagnetic or strong interactions, although they can have a narrow width because 
the mass-splittings in the spectrum are in general much smaller than the mass of the ground-state $B$-meson they ultimately decay to.
The corresponding lifetimes are of the order of $10^{-17}$ seconds or less and they typically do not live long enough to directly
experience a weak disintegration induced by the $b$-quark flavor transition. However, with the high luminosities achieved at the 
$e^+e^-$ colliders~\cite{Bevan:2014iga} and high production rates of $b\bar b$ pairs at the LHC, which already allow for sensitivities to branching
fractions at the level of $\sim10^{-10}$~\cite{Khachatryan2015}, some of these modes could become accessible to detection and investigation.

Of particular interest is the $B^*$ which is the partner of the $B$ in the heavy-meson doublet of the
$(b\bar q)$ system~\cite{Isgur:1991wq}. The $B^*$ are vectors and their $\Delta B=1$ decays have different sensitivities
to the short-distance structure of the transition as compared to those of the pseudoscalar $B$ mesons. Moreover, 
the hadronic matrix elements of these two mesons, which give 
the long-distance contributions to their decays, are related by heavy-quark symmetry. Thus, the interplay between $B$
and $B^*$ decays could prove useful in studies to test the standard model (SM) and search for new-physics (NP). This has 
actual and immediate interest as various anomalies have been detected in different charged- and neutral-current $B$ decays.
For instance, there is a long-standing discrepancy between the inclusive and exclusive determinations 
of the Cabibbo-Kobayashi-Maskawa (CKM) matrix element $V_{ub}$~\cite{Amhis:2014hma}
or tensions between the SM predictions and the measured $B\to D^{(*)}\tau\nu$ decay 
rates~\cite{Matyja:2007kt,Lees:2012xj,Lees:2013uzd,Aaij:2015yra}. These could be explained by 
NP altering the $V-A$ structure of the charged-current interaction characteristic of the SM~\cite{Crivellin:2009sd,Buras:2010pz,Crivellin:2014zpa,Aaij:2015bfa,
Crivellin:2009sd,Buras:2010pz,Crivellin:2014zpa,Aaij:2015bfa,Fajfer:2012vx,Becirevic:2012jf,Tanaka2012,Freytsis:2015qca}.

Moreover, in the course of the analyses done over run I, the LHCb experiment has reported a series of anomalies in various
(neutral-current) rare $b\to s\ell\ell$ decays~\cite{Aaij:2013pta,Aaij:2013qta,Aaij:2013aln,Aaij:2014pli,Aaij2015ag}
including a signal of lepton-universality violation~\cite{Aaij:2014ora}; remarkably, they can be largely accommodated
by a NP contribution to low-energy ``semileptonic'' operators selectively coupled to the muons, of the 
type~\cite{Descotes-Genon:2013wba,Altmannshofer:2013foa,Beaujean:2013soa,Horgan:2013pva,Alonso:2014csa,Hiller:2014yaa,Ghosh:2014awa},
\begin{eqnarray}
\mathcal{O}_9^\mu=\frac{\alpha_{\rm em}}{4\pi}(\bar s\gamma^\mu P_L b)\;(\bar\mu\gamma^\mu\mu),\hspace{1cm} 
\mathcal{O}_{10}^\mu=\frac{\alpha_{\rm em}}{4\pi}(\bar s\gamma^\mu P_L b)\;(\bar\mu\gamma^\mu\gamma_5\mu). \label{eq:semileptonic0}
\end{eqnarray}
More specifically, global fits to the $b\to s\ell\ell$ data point to scenarios where the NP contribution
to their respective Wilson coefficients are $C^{\mu,\,\rm NP}_9\simeq-1$ or $
C^{\mu,\,\rm NP}_9=-C^{\mu,\,\rm NP}_{10}\simeq-0.5$~\cite{Descotes-Genon:2013wba,Altmannshofer:2013foa,Beaujean:2013soa,Hurth:2014vma,
Altmannshofer:2014rta,Jager:2014rwa,Altmannshofer:2015sma}. These anomalies suggest the presence of new degrees of freedom
with non-universal lepton couplings and with masses in the TeV range that could be accessible to the direct searches at the 
LHC~\cite{Altmannshofer:2013foa,Altmannshofer:2014cfa,Hiller:2014yaa,Gripaios:2014tna,Sahoo:2015wya,Crivellin:2015mga,
Crivellin:2015lwa,Varzielas:2015iva,Sierra:2015fma,Celis:2015ara,Boucenna:2015raa,Crivellin:2015era,Becirevic:2015asa,
Lee:2015qra,Alonso:2015sja,Greljo:2015mma,Calibbi:2015kma,Altmannshofer:2015mqa}. 

Unfortunately, the interpretation of weak hadron decays is often obscured by the presence of long-distance
QCD effects whose impact in the analyses needs to be carefully 
assessed. This is specially true for some of the $b\to s\ell\ell$ anomalies which are found in observables
constructed from the decay rates of the semileptonic processes $B\to K^{(*)}\mu\mu$~\cite{Aaij:2013pta,
Aaij:2013qta,Aaij:2013aln,Aaij:2014ora,Aaij:2014pli} or $B_s\to\phi\mu\mu$~\cite{Aaij2015ag}. 
On one hand there are the hadronic matrix elements of local operators that can be parameterized in terms of 
functions of the invariant squared dilepton mass $q^2$ or form factors and whose description relies on the
accuracy of different nonperturbative methods~\cite{Isgur:1990kf,Burdman:1992hg,Manohar:2000dt,Beneke:2000wa,
Grinstein:2002cz,Jager:2012uw,Horgan:2013pva,Horgan:2013hoa,Descotes-Genon:2014uoa,Jager:2014rwa,Straub:2015ica}. On the other,
one needs to take into account the ``current-current'' four-quark operators, $\mathcal{O}_1$
and $\mathcal{O}_2$~\cite{Grinstein:1988me,Buchalla:1995vs,Chetyrkin:1996vx}, which in the SM stem from
the tree-level decay $b\to s c \bar c$. Hence, they come accompanied by large Wilson coefficients, $C_1$ and $C_2$,
and are not suppressed by either mixing angles or loop factors with respect to the contributions of the semileptonic operators.
They contribute to the neutral-current semileptonic decay amplitudes through an operator of the type,
\begin{equation}
\mathcal{T}_i^\mu(q^2)=i\int d^4x\,e^{i\,q\cdot x} T\,\{\mathcal{O}_i(0)\;,j_{\rm em}^\mu(x)\}, \label{eq:nonlocal0}
\end{equation}
produced by the contraction of the $c\bar c$ pair with the electromagnetic current and the off-shell photon eventually 
decaying into the dilepton pair. The hadronic matrix element
of these nonlocal operators receives dominant contributions from long-distance fluctuations of the charm-quark fields
manifested as charmonium resonances above the $c\bar c$ threshold. 

At high $q^2$ one can analytically continue eq.~(\ref{eq:nonlocal0}) into
the complex $q^2$-plane to perform an operator product expansion (OPE) which accurately describes
it in terms of a series of matrix elements of local operators matched 
perturbatively to QCD~\cite{Chay:1990da,Chibisov:1996wf,Grinstein:2004vb}. Continuing the
result back to the real $q^2$ gives the physical rates. This is called ``quark-hadron duality'' and its validity is
justified if $q^2$ is large and above the resonant contributions. More care is required when using the OPE
in a region with resonances where the violations to quark-hadron duality can be difficult 
to estimate. This is the case for the $b\to s\ell\ell$ exclusive decays 
since they are restricted to a region $q^2\leq(m_B-m_K)^2\lesssim22$ GeV$^2$ while the heaviest
charmonium state known is the $X(4660)$~\cite{Beylich:2011aq,Aaij:2013pta,Lyon:2014hpa}. 

In light of these difficulties, it is desirable to have alternative, theoretically cleaner processes probing the 
semileptonic operators in eq.~(\ref{eq:semileptonic0}) to confirm or to unambiguously characterize
the putative NP effect. A paradigmatic example is the $B_{d,s}\to\ell\ell$ decay, which depends on only
one hadronic quantity, a decay constant, that has been accurately determined using lattice simulations~\cite{Aoki:2013ldr}. 
The contribution to the amplitude of $C_9^\ell$, together with those of $C_1$ and $C_2$, vanish
due to the conservation of the vector current and the decay rate becomes sensitive only to $C_{10}^\ell$. 

In this paper we investigate the purely leptonic decays of the $B^*$ which are a particularly 
interesting class of decays. In contrast to those of their pseudoscalar siblings, with the $B^*$ being vector 
they are not chirally suppressed. This partly compensates for the shorter lifetime of the 
$B^*$ and makes them interesting to probe the short-distance structure of the muonic and electronic decays, 
specially in search for lepton-universality violation effects. They only depend on decay constants, 
which are calculable functions of the one of the $B$ in the heavy-quark limit and can be
accurately computed on the lattice. Interestingly, the neutral-current decay $B_s^*\to\ell\ell$ becomes 
sensitive to $C_9^\ell$ while the kinematics of the decay are such that $q^2\simeq28$ GeV$^2$, which
is well above the region of the charmonium resonances and the quark-hadron duality-violation to 
the contributions from $C_1$ and $C_2$ is expected to be much less of a concern. We will discuss
the prospects for producing this decay mode at the LHC and will also explore the possibility of
measuring the $B_s^*\to\ell\ell$ amplitude, instead, through scattering experiments at the $B_s^*$
resonance peak. We finish discussing also the charged-current $B_{u,c}^*\to\ell\nu$ decays which
can provide complementary information on the $b\to u\ell\nu$ and $b\to c\ell\nu$ transitions.

\section{The $B_s^*\to\ell\ell$ decay}

\subsection{Anatomy of the decay amplitude}

The state we are interested in is the partner of the $B_s$ in the ground-state spin doublet
of $(b\bar s)$ mesons. Its quantum numbers are $J^{PC}=1^{--}$, with a mass
$m_{B_s^*}=5415.4^{+2.4}_{-2.1}$ MeV~\cite{Agashe:2014kda} and a width that is experimentally unknown
although estimated to be of the order of 0.1 KeV (see Appendix).
In the SM, and neglecting electromagnetic corrections, the amplitude of the decay of a $B_s^*$ into a dilepton pair is:
\begin{align}
\mathcal{M}_{\ell\ell}&=\frac{G_F}{2\sqrt{2}}\lambda_{ts}\frac{\alpha_{\rm em}}{\pi}\Big[\left(m_{B_s^*}\,f_{B_s^*}\,C_9+2\,f_{B_s^*}^T\,m_b\,C_7\right)\bar \ell\slasheps \ell
+f_{B_s^*}C_{10}\bar \ell\slasheps\gamma_5 \ell\nonumber\\
&-8\pi^2\frac{1}{q^2}\sum_{i=1}^{6,8}C_i\,\langle 0|\mathcal{T}_i^\mu(q^2)|B_s^*(q,\varepsilon)\rangle\;\bar \ell\gamma_\mu \ell\Big],
\label{eq:ampl0}
\end{align}
where $G_F$ is the Fermi constant, $\lambda_{ts}=V_{ts}^*V_{tb}$, $m_b(\mu)$ the running $b$-quark mass in the $\overline{MS}$ scheme and $\varepsilon$ is the
polarization vector of the $B_s^*$. Furthermore, $q^2=m_{B_s^*}^2$. The information on the
short distance structure of the $b\to s$ transition is carried by the (renormalization scale dependent) Wilson 
coefficients of the weak Hamiltonian for $\Delta B=1$ processes~\cite{Grinstein:1988me,Buchalla:1995vs,Chetyrkin:1996vx}.
In particular, $C_{9,10}$ are the ones related to the short-distance semileptonic operators, eq.~(\ref{eq:semileptonic0}), and
$C_7$ is the coefficient of the ``electromagnetic penguin operator''~\cite{Grinstein:1987vj}.
The operators in the second line of eq.~(\ref{eq:ampl0}), correspond to either the four-quark operators, including those of
the current-current, $\mathcal{O}_{1,2}$, and the ``QCD-penguins'', $\mathcal{O}_{3,\ldots,6}$,
or the``chromo-magnetic penguin operator'', $\mathcal{O}_{8}$.~\footnote{For the 
definitions of these operators in this paper we follow the notations and basis introduced in ref.~\cite{Chetyrkin:1996vx}.}

The nonperturbative contributions enter through two types of matrix elements. Those of the 
local operators $\mathcal{O}_{7,9,10}$ are described by two decay constants,
\begin{equation}
\langle 0|\bar s\gamma^\mu b|B_s^*(q,\varepsilon)\rangle=m_{B_s^*}\,f_{B_s^*}\,\varepsilon^\mu,\hspace{1cm} 
\langle 0|\bar s\sigma^{\mu\nu} b|B_s^*(q,\varepsilon)\rangle=-i\,f_{B_s^*}^T(q^\mu\varepsilon^\nu-\varepsilon^\mu q^\nu),
\end{equation}
where $f_{B_s^*}^T(\mu)$ depends on the renormalization scale. In the heavy-quark limit, these are
related to the decay constant of the $B_s$~\cite{Manohar:2000dt},
\begin{eqnarray}
f_{B_s^*}=\,f_{B_s}\left(1-\frac{2\alpha_s}{3\pi}\right),\hspace{1cm}
f_{B_s^*}^T=f_{B_s} \left[1+\frac{2\alpha_s}{3\pi}\left(\log\left(\frac{m_b}{\mu}\right)-1\right)\right], \label{eq:fBsHQ}
\end{eqnarray}
where $\langle 0|\bar s\gamma^\mu \gamma_5 b|B_s(q)\rangle=-if_{B_s}\,q^\mu$ and we have neglected $\mathcal{O}(\alpha_s^2)$ corrections.

The second type of hadronic contribution enters, in the second line of eq.~(\ref{eq:ampl0}), through
the matrix element of the operator in eq.~(\ref{eq:nonlocal0}), induced by all the four-quark and the
chromomagnetic operators. At high $q^2\sim m_b^2$, one can exploit the hierarchy of scales
$\Lambda_{\rm QCD}\ll m_c\ll \sqrt{q^2}\sim m_b$ to expand this intrinsically nonlocal
object into a series of local operators matched perturbatively to QCD~\cite{Grinstein:2004vb}. The two
leading operators of the resulting OPE are equivalent to $\mathcal{O}_{7}$ and $\mathcal{O}_9$ so
that their matrix elements are described by the very same nonperturbative quantities
$f_{B_s^*}$ and $f_{B_s^*}^T$. In other words, the leading effect in the OPE is implemented by the redefinitions 
$C_7(\mu)\to C_7^{\rm eff}(\mu,\,q^2)$ and $C_9(\mu)\to C_9^{\rm eff}(\mu,\,q^2)$, where
the expressions of the matching are known up to next-to-leading order in $\alpha_s$~\cite{Seidel:2004jh,Grinstein:2004vb,Greub:2008cy}. 

A remarkable feature of this OPE is that the subleading operators in the expansion are 
suppressed by either $\mathcal{O}(\alpha_s\times \Lambda_{\rm QCD}/m_b)$ or $\mathcal{O}(\Lambda_{\rm QCD}^2/m_b^2)$~\cite{Grinstein:2004vb,Beylich:2011aq}
and are numerically small~\cite{Beylich:2011aq}. Nevertheless, one needs to remember
that the OPE is formally performed in the complex $q^2$ plane, away from the physical cuts and singularities~\cite{Chay:1990da,Chibisov:1996wf,Grinstein:2004vb};
there are the quark-hadron duality violations, not captured by the OPE to any order of $\alpha_s$ or $\Lambda_{\rm QCD}/m_b$ and known to appear in the
analytic continuation to the physical region. These are not understood
from first principles although it is believed they give rise to the oscillations characteristic of the resonances and to decrease 
exponentially into the higher $q^2$ region~\cite{Chibisov:1996wf}. For the kinematics of the $B_s^*$ decay, $q^2=m_{B_s^*}^2$ is well above
the charmonium states (and far below the bottomonium states) where local quark-hadron duality is expected to apply. 

\subsection{Numerical analysis}

The $B_s^*\to\ell\ell$ decay rate in the SM is then:
\begin{eqnarray}
\Gamma_{\ell\ell}=\frac{G_F^2\,|\lambda_{ts}|^2\alpha_{\rm em}^2}{96\pi^3}m_{B_s^*}^3f_{B_s^*}^2\left(|C_9^{\rm eff}(m_{B_s^*}^2)+
2\frac{m_b\,f_{B_s^*}^T}{m_{B_s^*}\,f_{B_s^*}}\,C_7^{\rm eff}(m_{B_s^*}^2)|^2+|C_{10}|^2\right), 
\end{eqnarray}
where we have neglected $\mathcal{O}(m_\ell^2/m_{B_s^*}^2)$ contributions.
For the implementation of the OPE in the present paper we follow~\cite{Grinstein:2004vb} and consider $m_c\ll m_b$, so that
an expansion up to $\mathcal{O}(m_c^2/m_b^2)$ is also implied. The relevant loop functions necessary for the matching at
$\mathcal{O}(\alpha_s$) are then obtained from refs.~\cite{Seidel:2004jh} and~\cite{Beneke:2001at}. For the running Wilson
coefficients $C_{1-8}$ of the weak Hamiltonian we use the next-to-leading log results, while for $C_{9,10}$ we include the
next-to-next-to-leading corrections calculated in~\cite{Gambino:2003zm}. The resulting renormalization scale dependence of the 
observables is very small, induced by either $C_9^{\rm eff}(\mu,\,q^2)$ at $\mathcal{O}(\alpha_s^2\times C_{1,2},\alpha_s\times C_{3-6})$ 
or by the combination $m_b(\mu)\,f^T_{B_s^*}(\mu)\,C_7^{\rm eff}(\mu,q^2)$ at $\mathcal{O}(\alpha_s^2)$~\cite{Grinstein:2004vb}.

\begin{table}[h]
\centering
\caption{Values for the relevant input parameters employed in the calculations
of this work. All are obtained from the PDG averages~\cite{Agashe:2014kda}, except
for $|\lambda_{ts}|$ which is determined from $|V_{cb}|$ and $|V^*_{tb} V_{ts}|/|V_{cb}|$ following ref.~\cite{Bobeth:2013uxa},
$f_{B_s}$, which is obtained from the $N_f=2+1$ FLAG average~\cite{Aoki:2013ldr} and $f_{B_s^*}/f_{B_s}$ which 
is taken from the HPQCD calculation in~\cite{Colquhoun:2015oha}.\label{tab:param}}
\begin{tabular}{|cc|cc|}
\hline
\multicolumn{2}{|c}{$G_F$}&\multicolumn{2}{c|}{$1.166 378 7(6)\times10^{-5}$ GeV$^{-2}$} \\
\hline
$m_b(m_b)$ &4.18(3)&$\alpha_s(m_Z)$&0.1184(7)\\
$m_c(m_c)$ & 1.275(25)&$\alpha_{\rm em}(m_b)$&1/134\\
$|\lambda_{ts}|$&0.0416(9)&$f_{B_s^*}/f_{B_s}$&0.953(23)\\
$m_{B_s^*}$&$5415.4^{+2.4}_{-2.1}$ MeV&$f_{B_s}$&227.7(4.5) MeV\\
\hline
\end{tabular}
\end{table}

In Tab.~\ref{tab:param} we show the values of the input parameters relevant for the numerical analysis
of this work. With these we obtain $C_9^{\rm eff}(m_b,m_{B_s^*}^2)=4.560+i\,0.612$ and $C_7^{\rm eff}(m_b,m_{B_s^*}^2)=-0.384-i\,0.111$.
For the decay constants, one can relate
them to $f_{B_s}$ using eqs.~(\ref{eq:fBsHQ}), which have been calculated accurately by different lattice
collaborations~\cite{Aoki:2013ldr}. One obtains $f_{B_s^*}/f_{B_s}=f_{B_s^*}^T(m_b)/f_{B_s}=0.95$.
Beyond the heavy-quark limit, the $f_{B_s^*}/f_{B_s}$ ratios have been calculated using QCD sum rules
~\cite{Neubert:1992fk,Huang:1995rv,Gelhausen:2013wia,Narison:2014ska,Lucha:2014hqa,Lucha:2015xua} and, 
recently, on the lattice by the HPQCD collaboration~\cite{Colquhoun:2015oha}. Interestingly, most of the QCD sum-rule
calculations obtain $f_{B_s^*}/f_{B_s}\simeq1.00-1.15>1$~\cite{Neubert:1992fk,Huang:1995rv,Gelhausen:2013wia}, 
while the latest sum-rule study~\cite{Lucha:2014hqa,Lucha:2015xua} and the lattice computation obtain a value that is
consistent with the one in the heavy-quark limit, \textit{viz.} $f_{B_s^*}/f_{B_s}=0.953(23)$~\cite{Colquhoun:2015oha}.
In this paper we will use this value as a benchmark for our predictions. It would be important, for 
further improvements of our analysis, to have independent calculations of this ratio. 
In addition, we are not aware of any computation of the tensor decay constant, $f_{B_s^*}^T(m_b)$. For this, we 
use the result given in the heavy-quark limit by the second equation in (\ref{eq:fBsHQ}) with an uncertainty $\mathcal{O}(\Lambda_{\rm QCD}/m_b)\sim 10\%$.

Our result for the decay rate then follows to be:
\begin{equation}
\Gamma_{\ell\ell}=1.12(5)(7)\times10^{-18} \;\;\text{GeV},\label{eq:Predllrate}
\end{equation}
where the first error stems from the one in the combination of CKM parameters $\lambda_{ts}$, 
and the second from the decay constants added in quadratures. The error from the residual renormalization
scale dependence is numerically very small, of the order of $1\%$ in the range from $\mu=m_b/2$ to $\mu=2 m_b$. 
Local quark-hadron duality violations to the OPE are difficult to estimate, especially because the kinematics
of the decay fall in a region where the data of $\sigma(e^+e^-\to\text{hadrons})$ is scarce. In any case, we observe that 
the few data points in the $\sqrt{s}\in[5,\,6]$ GeV region of this process are consistent with the result from QCD~\cite{Agashe:2014kda}. 
A better estimate could be obtained using the model of duality violation introduced in~\cite{Chibisov:1996wf}, fitted to the BES data on 
the $\sigma(e^+e^-\to\text{hadrons})$ across the charmonium region and adapted to the $b\to s\ell\ell$, as done in~\cite{Beylich:2011aq}. 
Extrapolating the results of this reference to the region $q^2\simeq m_{B_s^*}^2\pm2$ GeV$^2$ we observe that the duality-violating corrections to
$C_9^{\rm eff}$ are estimated to be less than a 1.5$\%$ of its short-distance contribution. 
 
\subsection{The branching fraction and prospects for experimental production}

The main difficulty for measuring this rare decay is that it has to compete with 
the dominant disintegration $B_s^*\to B_s\gamma$, which is an electromagnetic transition. The latter is suppressed
by a relatively small phase space~\footnote{The three-momentum of the recoiling particles is 
$|\vec{k}|=(m_{B_s^*}^2-m_{B_s}^2)/(2m_{B_s^*})=0.048$ GeV to be compared with the one of the leptonic rare decay, $|\vec{k}|\simeq m_{B_s^*}/2\simeq2.71$ GeV. 
Strong decays of the $B_s^*$ are forbidden by phase space.}
while the vector nature of the $B_s^*$ makes the former not to be chirally suppressed, as it is the case of $B_s\to\mu\mu$.     
In order to calculate the branching fraction and study the feasibility of measuring this decay mode
we thus need the $B_s^*$ width produced by the electromagnetic transition that is not measured and
theoretically not very well known. 

The $B_s^*\to B_s\gamma$ rate is determined by a hadronic transition magnetic moment, $\mu_{bs}$, 
that can be estimated using heavy-quark and chiral perturbation effective theories from the equivalent decays in the $D^{(*)}$
system~\cite{Cho:1992nt,Amundson:1992yp,Stewart:1998ke}. In the Appendix we define this quantity and show a determination along these
lines using current experimental data and recent lattice QCD results as input. We obtain $\Gamma=0.10(5)$ KeV, 
which is consistent with the results of the earlier analyses and different quark-model calculations
~\cite{Goity:2000dk,Ebert:2002xz,Choi:2007se}. The conclusions of our study are then hindered by this large uncertainty in $\Gamma$; 
it is important to stress, though, that this concerns a single hadronic quantity that can be calculated in the lattice as recently demonstrated for
the $D^{(*)}$ system in ref.~\cite{Becirevic:2009xp}. Progress in this direction is essential for a conclusive 
assessment on the interest of this mode and for the experimental prospects for its detection and measurement.  

With these caveats, we proceed to combine eq.~(\ref{eq:Predllrate}) with our estimate of $\Gamma$ and obtain
a branching fraction that in the SM is in the range:
\begin{eqnarray}
\mathcal{B}^{\rm SM}(B_s^*\to\ell\ell)=1.12(9)\times10^{-11}\,\left(\frac{0.10(5)\,\text{KeV}}{\Gamma}\right)=(0.7-2.2)\times10^{-11},\label{eq:PredllBR}
\end{eqnarray}
irrespective of the lepton flavor, and where we have added in quadratures the uncertainties in the $\Gamma_{\ell\ell}$ rate.
This is a very small branching fraction, lying an order of magnitude below $\mathcal{B}^{\rm SM}(B_d\to\mu\mu)$ \cite{Bobeth:2013uxa}
and the rarest decay ever detected in an experiment, $K\to\pi\nu\bar \nu$~\cite{Artamonov:2008qb}. 

Measuring $B_s^*\to\ell\ell$ is far from the reach of the Super $B$-factories, as for example, Belle~II expects to collect no more than $5\times 10^8$ of $B_s^*$ after 
5 ab$^{-1}$ at the $\Upsilon(5S)$~\cite{Bevan:2014iga}. On the other hand, and given the large production rates of $b\bar b$ pairs in high-energy
$pp$ collisions, it could be searched for at the LHC. To make an estimate, let us start with the 100 $B_s\to\mu\mu$ events expected  
after run I by the combined analysis of the LHCb (3 fb$^{-1}$) and CMS (25 fb$^{-1}$)~\cite{CMS:2014xfa}, while the full results from the ATLAS 
experiment have not been reported yet (see e.g.~\cite{Aad:2012pn}). In the course of runs II and III the experiments at CERN will collect
$\sim10$ times more data~\cite{Bediaga:1443882,CMS:2013vfa} and the $b\bar b$ production will be boosted further by a factor $\sim2$ by the higher 
cross section at $\sqrt{s}=14$ TeV; after the high-luminosity (HL-LHC) upgrade, a factor $\sim10$ more of data is expected~\cite{Bediaga:1443882,CMS:2013vfa}. 
Rescaling up naively the current
$B_s\to\mu\mu$ events, we estimate $\sim3\times10^3$ events by the end of run III and $\sim3\times10^4$ after the full run in the HL-LHC phase. 

In order to use this to estimate the number of $B_s^*\to\ell\ell$ we need to know the fraction of $B_s^*$ produced by the hadronization of the $b$-quark
as compared to the one for $B_s$ meson. In the heavy-quark limit this can be derived by simple helicity arguments that suggest that the $B_s^*$ are 
produced 3 times more often than the $B_s$~\cite{Manohar:2000dt}. This has been confirmed for the $B^*$ system in measurements at the $Z^0$ peak by 
LEP~\cite{Buskulic:1995mt}. For the $B_s^*$ system this factor is even larger in the production at the $\Upsilon(5S)$~\cite{Agashe:2014kda,Bevan:2014iga}. 
This means that most of the $B_s^0$ mesons detected at the LHC should have been produced through a $B_s^*\gamma$ decay. Taking
this into account and that the branching fraction of the $B_s\to\mu\mu$ is $\sim300$ times larger than (\ref{eq:PredllBR}) we finally estimate that of
the order of 10 (100) $B_s^*\to\ell\ell$ decays could be produced by the end of the run III (HL-LHC). Whether or not this could be measured by the
LHC experiments will depend on a careful assessment of the backgrounds, but in general, we would expect the signal to manifest as a separate peak
to the right of the $B_s$ distribution in the invariant dilepton mass of the $B_{d,s}\to\mu\mu$ measurements. The estimate for
$B_s^*\to ee$ differs from the previous one because of the different detector efficiencies for muons and electrons. Interestingly, the 
electronic mode has no background from the $B_s\to ee$ decay since this mode is very suppressed.

\subsection{Resonant $B_s^*$ production in $\ell^+\ell^-$ scattering}

We speculate here about a completely different experiment to measure the $B_s^*\to\ell\ell$ 
rate and we briefly study its feasibility. It consists of producing a $B_s^*$ through resonant 
$\ell^+\ell^-$ scattering, where the $\ell$ could either be an electron or a muon. The idea is that
the loop- and CKM-suppression of the amplitude is largely compensated by the resonant
enhancement in the cross-section from the small width of the $B_s^*$. Moreover, we expect that the production of a single 
$b$ or $\bar b$ quark at $\sqrt{s}\sim5.5$ GeV from a $\ell^+ \ell^-$ collision would give such a distinct
experimental signature that it could be easily disentangled from other electromagnetically produced 
$\ell^+\ell^-\to\text{hadron}$ events.

A calculation of the cross section of $\ell^+\ell^-\to B_s^*\to B_s\gamma$ and its charged-conjugate 
(we omit $CP$-violation effects) gives:
\begin{eqnarray}
\sigma(s)=\frac{24\pi\,m_{B_s^*}^2}{s}\left(\frac{s-m_{B_s}^2}{m_{B_s^*}^2-m_{B_s}^2}\right)^3\frac{\Gamma_{\ell\ell}\Gamma}{(s-m_{B_s^*}^2)^2+m_{B_s^*}^2\Gamma^2},\label{eq:crosssec0}
\end{eqnarray}
where we have assumed $s\simeq m_{B_s^*}^2$ so that the rates $\Gamma_{\ell\ell}$ and $\Gamma$
are evaluated for the $B_s^*$ on-shell and have neglected lepton mass effects and  non-resonant contributions
to the process. It follows that:
\begin{eqnarray}
\sigma_0=\sigma(m_{B_s^*}^2)=\frac{24\pi}{m_{B_s^*}^2}\mathcal{B}(B_s^*\to\ell\ell)\label{eq:crosssec1}
\end{eqnarray}
and using the results in eq.~(\ref{eq:PredllBR}), we obtain:
\begin{eqnarray}
\sigma_0=(7-22)\,\text{fb},\label{eq:crosssec2}
\end{eqnarray}
where the large error originates again from $\Gamma$. This is a small cross section,characteristic of
other weak processes like neutrino-nucleon scattering which occurs at $\sigma_{\nu N}\sim1-10$~fb.

In order to assess if this process is accessible to experimental study, we need to consider the fact that
the energy of the particles in the beams distribute over certain range whose size is quantified by the ``energy spread''
of the accelerator, $\Delta E$. For current $e^+e^-$ colliders, and for the center-of-mass energies under
consideration, $\Delta E_e\sim 1$ MeV, which is much larger than $\Gamma$ so that only a small fraction of the collisions would occur where the
cross-section is maximal. A better control over the energy spread could be achieved at a $\mu^+\mu^-$ collider,
although the minimum that has been projected for such hypothetical facility is $\Delta E_\mu\sim100$ KeV for
the energies of interest~\cite{Delahaye:2013jla}, which is also much larger than $\Gamma$.   

For the sake of simplicity let us assume that the energy of the particles in the colliding beams 
spreads uniformly within the interval $m_{B_s^*}/2\pm\Delta E$, and that $\Gamma\ll\Delta E$. In this case, 
the average cross-section $\bar \sigma$ is:
\begin{equation}
\bar \sigma=\frac{6\pi^2}{m_{B_s^*}^2}\frac{\Gamma_{\ell\ell}}{\Delta E}=\frac{\pi}{4}\frac{\Gamma}{\Delta E}\sigma_0,\label{eq:crosssec3}
\end{equation}
and using the $\sigma_0$ and the $\Delta E$ discussed above, $\bar \sigma\sim 1$~ab and
$\bar \sigma\sim 10$~ab for the $e^+e^-$ and $\mu^+\mu^-$ colliders, respectively. Producing these processes
experimentally might be at reach in the future as, for example, SuperKEKB expects to produce more than
10 ab$^{-1}$/yr of $e^+e^-$ collisions within the next decade~\cite{Bevan:2014iga}.

Another interesting possibility is considering the orbital excitations of the $B_s$,
in particular the lighter mesons in which the $s$-quark is in the $P$-wave orbital. This corresponds
to two almost degenerate heavy-quark doublets which are predicted to have the quantum numbers 
$J^{P}=(0^{+},\,1^{+})$ and $(1^{+},\,2^{+})$, masses of the order of $\sim5.8$ GeV and narrow
widths, $\Gamma\sim0.01-1$ MeV~\cite{Godfrey:1986wj,Isgur:1991wq,Eichten:1993ub,Falk:1995th,Ebert:1997nk,DiPierro:2001uu,
Bardeen:2003kt,Vijande:2007ke,Gregory:2010gm,Colangelo:2012xi,Bernardoni:2015nqa}. 
The last pair of axial-vector and tensor states have been identified with the two observed $B_{s1}(5830)$ and $B_{s2}^*(5840)$ 
resonances~\cite{Abazov:2007af,Aaltonen:2007ah,Aaij:2012uva}, where the width of $B_{s2}^*$ has also been measured,
$\Gamma(B_{s2}^*)=1.56(13)(47)$ MeV~\cite{Aaij:2012uva}.  

These resonances could be produced in resonant $\ell^+\ell^-$ scattering. Their widths are closer to the energy spreads
achievable in current and projected accelerators and the scattering would enjoy more luminosity over the resonance region,
albeit at the cost of a reduction of the resonant enhancement of the cross section. If the leptonic weak rates for these states were of
the same order of magnitude as $\Gamma_{\ell\ell}$, one can see from eq.~(\ref{eq:crosssec0}) and (\ref{eq:crosssec1}),
that the cross-section for the production would scale as $\Gamma/\Gamma^*$, where $\Gamma^*$ is the corresponding width.
Besides this, studying the leptonic rates and amplitudes for the orbital excitations is interesting because 
their quantum numbers lead to different independent sensitivities to the short-distance structure of the $b\to s\ell\ell$ weak
transition. They are theoretically clean processes provided the relevant decay constants can be calculated
accurately and their widths determined. The fact that these states are quite heavier than the $B_s^{(*)}$ 
could also allow for studying the validity of quark-hadron duality in more detail.

\section{The $B^{*-}\to\ell^-\bar\nu$ decays}

The idea of studying the weak disintegrations of the unstable heavy-light systems can be straightforwardly 
applied to the charged-current leptonic decays of the excited $B_{i}^{*\pm}$ states, where $i=u,\,c$. Similarly to the $B_s^*$, 
the vector nature of these resonances partly compensates for the 
shorter life-time as compared to the same decays of their pseudoscalar partners. 
Nevertheless, $B_i\to\ell\nu$ is difficult to observe not only because of the
chiral suppression of the rates in the SM but also because the neutrino in the final state. Only
the $B_u\to\tau\nu$ has been detected~\cite{Lees:2012ju,Adachi:2012mm} while limits at the level of
$\mathcal{B}<10^{-6}\,(95\%\,\text{C.L.})$ have been placed on the electronic and muonic modes~\cite{Satoyama:2006xn}.
The decay channels of the $B_c$ remain unmeasured to a large extent but important progress is expected at
the LHC~\cite{Gouz:2002kk,Anderlini:2014dha}. 

The complementarity between the decays of the purely leptonic decays of the $B_{i}$ and $B_{i}^*$ can be explored
by modifying the characteristic charged-current $V-A$ interaction of the SM as,
\begin{equation}
\mathcal{L_{\rm c.c.}}=-\frac{4G_F}{\sqrt{2}}V_{ib}\left((1+\epsilon_L^{i\ell})(\bar u_i\gamma^\mu P_L b)\,(\bar\ell\gamma_\mu P_L\nu)+
\epsilon_R^{i\ell}(\bar u_i\gamma^\mu P_R b)\,(\bar\ell\gamma_\mu P_L\nu) \right),
\end{equation}
where $\epsilon_{L,R}^{\ell i}$ are Wilson coefficients encoding NP left-handed or right-handed currents that
could be lepton dependent. Contributions of this type are among the possible explanations for the different anomalies 
found in the $b\to u\ell\nu$ and $b\to c\ell\nu$ transitions~\cite{Crivellin:2009sd,Buras:2010pz,Crivellin:2014zpa,
Aaij:2015bfa,Fajfer:2012vx,Becirevic:2012jf,Tanaka2012,Freytsis:2015qca}. The $B_i^{(*)}\to\ell\nu$ decay rates are:
\begin{eqnarray}
\Gamma_{\nu\ell}=\frac{G_F^2}{8\pi}|V_{ib}|^2(1+\epsilon_L^{i\ell}-\epsilon_R^{i\ell})^2m_{B_i}\,f_{B_i}^2 m_\ell^2,\hspace{0.5cm} \Gamma_{\nu\ell}^*=\frac{G_F^2}{12\pi}|V_{ib}|^2(1+\epsilon_L^{i\ell}+\epsilon_R^{i\ell})^2m_{B_i^*}^3\,f_{B_i^*}^2, 
\end{eqnarray}
where we neglect subleading $\mathcal{O}(m_\ell^2/m_{B_i^*}^2)$ corrections. The different 
quantum numbers of the $B_i$ and $B_i^*$ mesons make their amplitudes sensitive to different and orthogonal
combinations of the coefficients $\epsilon_{L,R}^{i\ell}$. This can be exploited better by 
looking at the ratio of branching fractions:
\begin{eqnarray}
R_{i\ell}^*=\frac{\mathcal{B}(B_i^*\to\ell\nu)}{\mathcal{B}(B_i\to\ell\nu)}=\frac{2}{3}\frac{m_{B_i^*}}{m_{B_i}}\left(\frac{f_{B_i^*}}{f_{B_i}}\right)^2 \frac{\tau_{B^*_i}}{\tau_{B_i}}
\left(\frac{m_{B_i^*}}{m_\ell}\right)^2(1+4\epsilon_R^{i\ell})+\mathcal{O}(\epsilon^{i\ell}_{L,R})^2,
\end{eqnarray}
which are clean observables sensitive to right-handed currents. 

\begin{table}[h]
\centering
\caption{Results for the branching fractions of the different charged-current leptonic $B_i^{(*)}$ decays
considered in this work ($i=u,\,c$ and $\ell=e,\,\mu$). The uncertainties in the $B_i^*$ decays are dominated by the error of their
widths whereas those of the $B_i$ are of a few percent relative to the central value.\label{tab:results}}
\begin{tabular}{|c|c|cc|}
\hline
&$B_i^*\to\ell\bar\nu$&$B_i\to e\bar\nu$&$B_i\to\mu\bar\nu$ \\
\hline
$i=u$&$0.6^{+0.3}_{-0.2}\times10^{-9}$&$1.2\times10^{-12}$&$4.9\times 10^{-7}$\\
$i=c$&$1.3^{+0.4}_{-0.2}\times10^{-5}$&$2.6\times10^{-9}$&$1.6\times10^{-4}$\\
\hline
\end{tabular}
\end{table}

In order to know the practical interest of these modes we need to know the width (or lifetime) of the $B_i^*$ mesons,
induced by their electromagnetic decay, which can be again estimated as explained in the Appendix, giving $\Gamma_u=0.50(25)$ KeV and $\Gamma_c=0.030(7)$ KeV.
In Table~\ref{tab:results} we show the subsequent predictions for the branching fractions of the $B_i^*\to\ell\nu$ decays as compared to the
electronic and muonic modes of the decays of the $B_i$ mesons.~\footnote{For the masses of the $B_i$ and $B^{*}_i$ we take the PDG 
averages~\cite{Agashe:2014kda} and we use lattice calculations for the rest of the inputs, $f_B=190.5(4.2)$ MeV~\cite{Aoki:2013ldr}, 
$f_B^*/f_B=0.941(26)$~\cite{Colquhoun:2015oha}, $m_{B_c^*}=6315(8)$ MeV, $f_{B_c}=489$ MeV~\cite{Chiu:2007km} and assume $f_{B_c}^*/f_{B_c}=1$. For the
CKM matrix elements we use the inclusive determinations $|V_{ub}|=4.13(49)\times10^{-3}$~\cite{Agashe:2014kda} and $|V_{cb}|=0.0424(9)$~\cite{Gambino:2013rza}.}  
We observe that for the  $B_u^*$ state the branching fraction is very small although it is larger than the one of the $B_u\to e\nu$ mode.
On the other hand, the branching fraction of the $B_c^*$ state is not unreasonably small. It is only an order of magnitude smaller than the
$B_c\to \mu\nu$ mode and still much larger than $B_c\to e\nu$. 

\section{Conclusions}

The vector $B^*$ states are very narrow resonances because of the phase-space suppression suffered by
their dominant electromagnetic decays. The fact that the purely leptonic decays of the $B^*$ are not chirally
suppressed compensates for their short lifetimes and the resulting branching fractions are not much smaller 
(for muons) or are even larger (for electrons) than those of the leptonic decays of the pseudoscalar $B$ mesons. 
The $B_s^*\to\ell\ell$ decay is especially interesting since it could provide a clean window to a  
class of semileptonic $b\to s\ell\ell$ operators, in particular $\mathcal{O}_9$, that 
could contain information of new physics at the TeV scales. 

The advantage of $B_s^*\to\ell\ell$ over other decays (e.g. semileptonic
rare decays) is its theoretical cleanness since $\textit{(i)}$ the amplitude only depends on decay constants which are 
determined accurately in the lattice; and $\textit{(ii)}$ the invariant mass of the process is well above
the charmonium resonances and the application of an operator-product expansion for the nonlocal contributions
of eq.~(\ref{eq:nonlocal0}) via quark-hadron duality (which always accompany the contributions of $\mathcal{O}_9$) 
is well justified.  

The $B_s^*\to\ell\ell$ decay rate can be accurately predicted in the standard model. Using some 
estimates for the unmeasured width of the $B_s^*$, we obtained that the branching fraction for this process is
$\sim10^{-11}$ which could be within reach in the next series of experiments at the LHC. 
More accurate determinations of the width, for example using lattice techniques, are important
since this remains the major obstacle for an accurate calculation of the branching fraction of the decay.

The same amplitudes can be measured using a different strategy based
on resonant $\ell^+\ell^-\to B_s^*\to B_s\gamma$ scattering. The idea is that the strong suppression of the 
amplitude is compensated by a large enhancement from the small width of the resonance. In fact, the cross-section
at the mass of the $B_s^*$ is of the same order of magnitude as, for example, the one
for neutrino-nucleon scattering. Taking into account the energy spread of the beams reduces 
the effective cross-section and we estimated that this would be of the order of $1-10$ ab for the current
or projected accelerators. Other orbitally excited $(b\bar s)$ states are also interesting as they have broader
widths and can present different sensitivities to the same underlying effective operators.

The same type of analysis can be extended to the leptonic charged-current decays of the $B^{*\pm}_{u,c}$ mesons studying
their complementarity with those of the $B^{\pm}_{u,c}$ mesons. For instance, the 
sensitivities induced by their different quantum numbers could be used to test for the left-handedness of the charged-current 
transitions. For the  $B_u^*$ state the branching fraction results to be $\sim10^{-9}$ whereas for the $B_c^*$ state is 
of the order of $10^{-5}$ and only an order of magnitude smaller than the $B_c\to \mu\nu$ mode. 

Notwithstanding the entertainment value (at least to the present authors) of this investigation,
we maintain that the rates of the proposed experiments are not ridiculously small; the creativity
and prowess of the experimenter should not be discounted.~\footnote{During a seminar right after
publication of~\cite{Grinstein:1988me} one of the authors was  ridiculed  for the preposterous
notion that $B\to X_s\ell\ell$ would ever be measured.}

\section{Acknowledgments}

We thank U.~Egede for encouragement and discussions. This work was supported 
in part by DOE grant DE-SC0009919. JMC has received funding 
from the People Programme (Marie Curie Actions) of the European Union's 
Seventh Framework Programme (FP7/2007-2013) under REA grant agreement
n PIOF-GA-2012-330458 and acknowledges the Spanish Ministerio de Econom\'ia y 
Competitividad and european FEDER funds under the contract FIS2011-28853-C02-01
for support. 

\appendix

\section{$B_s^*\to B_s\gamma$ and the $B_s^*$ width}

The $B_s^*\to B_s\gamma$ decay rate (and  $B_s^*$ width) can be estimated in a 
model-independent way using heavy-quark and chiral effective theories~\cite{Cho:1992nt,Amundson:1992yp,Stewart:1998ke}. 
The amplitude of this transition is:
\begin{equation}
\mathcal{M}_\gamma=\langle B_s(q-k)|j^\mu_{\rm e.m.}|B_s^*(q,\,\varepsilon)\rangle\eta^*_\mu=
e\,\mu_{bs}\,\epsilon^{\mu\nu\rho\sigma}\eta_\mu^*q_\nu k_\rho\varepsilon_\sigma, \label{eq:emamp}
\end{equation}
where $e$ is the electric charge, $\eta$ ($k$) and $\varepsilon$ ($q$) are the polarization vectors 
(four-momenta) of the photon and the $B_s^*$ respectively, and with $\mu_{bs}$ a nonperturbative
magnetic moment. The electromagnetic decay rate then follows as: 
\begin{equation}
\Gamma_\gamma=\frac{\alpha_{\rm em}}{3}\mu_{bs}^2\,|\vec{k}|^3. \label{eq:emrate}
\end{equation}
The magnetic moment can be separated into two components, 
$\mu_{bs}=\mu_b+\mu_s$. The first one, $\mu_b$, is obtained in heavy-quark effective theory simply as the
magnetic moment of the $b$-quark appearing in the effective Lagrangian at $\mathcal{O}(\Lambda_{\rm QCD}/m_b)$~\cite{Cho:1992nt}: 
\begin{equation}
 \mu_b=-\frac{1}{3\,m_b}.\label{eq:muheavy}
\end{equation}
The light component, $\mu_s$, involves the long-distance, heavy-quark spin conserving contributions
of the light-quarks which are described by pion and kaon fluctuations coupled 
to the heavy hadron within the framework of chiral perturbation theory~\cite{Amundson:1992yp}. At 
leading order in the chiral expansion, $\mu_s$, $\mu_u$ and $\mu_d$ are related by $SU(3)$
flavor symmetry to a single nonperturbative parameter, $\mu_l=Q_l\beta$ where $Q_l$ is 
the electric charge of the light quark $l$. Leading $SU(3)$-breaking corrections are given at 
the next-to-leading order by pion and kaon loops~\cite{Amundson:1992yp}:
\begin{eqnarray}
\delta\mu_u=-\frac{g_1^2\,m_\pi}{4\pi f_\pi^2}-\frac{g_1^2\,m_K}{4\pi f_\pi^2},\;\;\delta\mu_d=\frac{g_1^2\,m_\pi}{4\pi f_\pi^2}, \;\;\delta\mu_s=\frac{g_1^2\,m_K}{4\pi f_\pi^2}, \label{eq:mulight}
\end{eqnarray}
where $f_\pi\simeq131$ MeV is the pion semileptonic decay constant, and
$m_\pi\simeq139$ MeV and $m_K\simeq495$ MeV are the meson masses. The $g_1$ is the effective
coupling of the pseudoscalar and heavy mesons and which has been obtained from lattice calculations, $g_1\simeq 0.50$~\cite{Becirevic:2009yb,Detmold:2012ge,Bernardoni:2014kla}.  

In the heavy quark-limit $\beta$ relates the magnetic transitions of the $B^*$ mesons
to those in the charm sector, where experimental information is available. In particular:
\begin{equation}
\Gamma(D^{*\pm}\to D^\pm\gamma)=\Gamma(D^{*\pm})\times\mathcal{B}(D^{*\pm}\to D^{\pm}\gamma)=1.33(33)\;\;\text{KeV}, 
\end{equation}
where we have used the results obtained by BaBar~\cite{Lees:2013zna} on the $D^{*\pm}$ width and the CLEO results
for the branching fraction~\cite{Bartelt:1997yu}. Equating eq.~(\ref{eq:emrate}) to this experimental result one obtains that
$\mu_{cd}=-0.46(5)$ GeV$^{-1}$, and then, $\beta=3.41(16)$ GeV$^{-1}$. This value for $\mu_{cd}$ together with the one 
readily obtained as a prediction for $\mu_{cu}=2.1(1)$ GeV$^{-1}$, compare well with the results of the recent lattice
calculation of ref.~\cite{Becirevic:2009xp}, $\mu_{cd}=-0.2(3)$ GeV$^{-1}$ and $\mu_{cu}=2.0(6)$ GeV$^{-1}$. With this, we are now
ready to predict $\mu_{bs}=-0.64(5)$ GeV$^{-1}$, which translates into $\Gamma_\gamma\simeq0.11(1)$ KeV.

This value is consistent with the predictions obtained using a similar formalism and older 
data~\cite{Cho:1992nt,Amundson:1992yp,Stewart:1998ke} and with those in various quark models~\cite{Goity:2000dk,Ebert:2002xz,Choi:2007se}.
Nevertheless, and beyond the experimental uncertainties, we expect this result to receive sizable 
corrections from the chiral expansion or breaking the heavy-quark symmetry (e.g. the heavy-quark dependence
of the constant $\beta$ or through recoil corrections of the heavy mesons). These scale like $\mathcal{O}(m_K^2/\Lambda_{\rm\chi SB}^2)$ and 
$\mathcal{O}(\Lambda_{\rm QCD}/m_c)$ respectively, each of which could be as large as a 25\% correction. A manifestation of this problem
is the large size of the kaon loops (about 1/2 of the total contribution) which makes our results very sensitivity to the exact value of
$g_1$~\cite{Becirevic:2009yb,Detmold:2012ge,Bernardoni:2014kla}, or to whether one implements higher $SU(3)$-breaking corrections
phenomenologically by using $f_K=1.22 f_\pi$ in the kaon loops~\cite{Amundson:1992yp} or not. With all this in mind, we will use,
\begin{equation}
\Gamma_\gamma=0.10(5)\;\;\text{KeV}, 
\end{equation} 
for the phenomenological discussion of this paper.

This analysis can be extended to calculate the electromagnetic decay rates of the $B_{u,c}^{*\pm}$ mesons. In case of
the $B_u^*$ one simply implements the contribution of the light quark $\mu_u=2/3 \,\beta+\delta\mu_u$, giving $\Gamma_\gamma=0.50(25)$~KeV. 
For the case of the $B_c^*$, the contributions from the two heavy quarks are of the type induced by eq.~(\ref{eq:muheavy}),
leading to $\Gamma_\gamma=0.030(7)$~KeV, where the error has been estimated as of $\mathcal{O}(\Lambda_{\rm QCD}/m_c)$.

\end{document}